\documentclass[12pt]{article}   
\usepackage{epsfig}
\newcommand{\mysection}{\setcounter{equation}{0}\section}

\def\beq{\begin{equation}}
\def\eeq{\end{equation}}
\def\beqa{\begin{eqnarray}}
\def\eeqa{\end{eqnarray}}

\newlength{\dinwidth} \newlength{\dinmargin}
\begin{document}

\begin{center}
{\Large \bf The theoretical top quark cross section at the Tevatron and 
the LHC}
\end{center}
\vspace{2mm}
\begin{center}
{\large Nikolaos Kidonakis$^{(1)}$ and Ramona Vogt$^{(2,3)}$}\\
\vspace{2mm}
{}$^{(1)}${\it Kennesaw State University, Physics \#1202\\
1000 Chastain Rd., Kennesaw, GA 30144-5591, USA}\\
{}$^{(2)}${\it Lawrence Livermore National Laboratory\\
Livermore, CA 94551, USA}\\
{}$^{(3)}${\it Physics Department, University of California at Davis\\
Davis, CA 95616, USA}
\end{center}

\begin{abstract}
We present results for the top quark pair cross section at the Tevatron 
and the LHC. We use the resummed double differential cross 
section, employing the fully kinematics-dependent soft anomalous dimension 
matrices, to calculate the soft-gluon contributions at next-to-next-to-leading 
order (NNLO).  We improve and update our previous estimates by refining our 
methods, including further subleading terms, and employing the most recent 
parton distribution function sets.  The NNLO soft corrections significantly 
enhance the NLO cross section while considerably reducing the scale 
dependence.  We provide a detailed discussion of all 
theoretical uncertainties in our calculation, including kinematics, 
scale, and parton distributions uncertainties and clarify the differences 
between our work and other approaches in the literature. 
\end{abstract}

\thispagestyle{empty} \newpage \setcounter{page}{2}

\mysection{Introduction}

The top quark holds a special place in the Standard Model of particle 
physics as the heaviest elementary particle. Since the discovery 
of the top quark via $t{\bar t}$ production in proton-antiproton collisions 
at the Tevatron \cite{CDFD0} its mass \cite{topmass} and production cross 
section \cite{CDFcs,D0cs} have been determined with increasing accuracy. 
There is also now evidence for 
single-top production at the Tevatron \cite{CDFD0st}. 
At the LHC,  both the
$t {\bar t}$ and single top production cross sections 
will be two orders of magnitude higher than at the Tevatron.
For recent reviews of top quark physics in hadron colliders 
see \cite{topreview}.

The top quark cross section receives large corrections from 
soft-gluon contributions near threshold which can be formally resummed. 
The resummation at next-to-leading logarithmic (NLL) 
accuracy for hard scattering cross sections and, in particular, top 
quark pair production was performed in Ref.~\cite{NKGS}. 
(For recent results on single top production, see Ref.~\cite{NKsingletop}). 
To achieve such accuracy, it is necessary to derive the soft anomalous 
dimension matrix, which controls noncollinear soft-gluon emission, to one 
loop. At NLL the color structure of the hard scattering enters in a non-trivial
way and each partonic process has to be treated separately.
The soft anomalous dimension matrix is dependent on all the 
kinematical variables.
Thus this is a fully differential calculation which can be applied to total 
cross sections as well as to differential cross sections, such as transverse 
momentum and rapidity distributions. 

Later, another formalism \cite{BCMN}
was proposed for the total cross section only. This calculationally 
simpler approach does not, however, involve the exact differential kinematics 
and instead makes the approximation that the NNLO and NLO rapidity dependence 
is the same.  Hence, numerical deviations from the exact kinematics-sensitive 
result can appear.  (For a detailed discussion and a numerical comparison 
in the context of direct photon production, see Ref.~\cite{GSWV}). 

In Refs.~\cite{NKGS,BCMN}, the resummation is performed in moment space.
Since the expression for the resummed cross section diverges at the 
Landau pole, a prescription is needed to define the physical resummed 
cross section when inverting from moment to momentum space. 
Alternatively, to avoid prescription 
dependence, the resummed cross section can be expanded to NNLO or higher orders.

The formalism of Ref.~\cite{NKGS} was used in detailed phenomenological 
studies \cite{NKtop,KLMV} where NNLO expansions were provided at NNLL 
accuracy, after matching with the complete NLO cross section. 
Results were provided in both single-particle-inclusive (1PI) and 
pair-invariant-mass (PIM) kinematics.  The kinematics ambiguity was found to be
an important source of uncertainty. 
In 1PI kinematics the soft logarithms are of the form $[\ln^k(s_4/m^2)/s_4]_+$ 
with $m$ the top quark mass and $s_4$ the sum of the Mandelstam 
invariants, $s_4=s+t_1+u_1$.
Near threshold, $s_4\rightarrow 0$. The soft-gluon corrections to the 
double differential cross section, $d^2\sigma/(dt_1 du_1)$, were calculated. 
In PIM kinematics, the soft logarithms are of the form $[\ln^k(1-z)/(1-z)]_+$ 
with $z=M^2/s$, where $M^2$ is the $t \overline t$ pair mass squared. 
Near threshold, $z\rightarrow 1$.  The soft gluon corrections to the 
double differential cross section, $d^2\sigma/(dM^2 d\cos\theta)$, where 
$\theta$ is the scattering angle in the partonic center-of-mass frame, 
were calculated.  The cross section in PIM kinematics was found to be smaller 
than the 1PI result.  The difference, an uncertainty due to 
uncalculated terms, was found to be larger than the scale variation. 
In Ref.~\cite{KLMV}, results were also given for the exact scale variation 
at NNLO. The magnitude of this variation was also found to depend on the 
kinematics.

The formalism of Ref.~\cite{BCMN} was also used to derive NLL resummed 
numerical results (recently updated \cite{CFMNR}) for top pair production  
at the Tevatron and LHC.  The corrections beyond NNLO are negligible \cite{BCMN}
(also shown to be small in Ref.~\cite{NKNNNLO}); hence the resummed cross 
section is numerically very similar to the NNLO expansion at the given 
logarithmic accuracy. 
However, a minimal prescription was used to define the resummed cross section 
in Ref.~\cite{BCMN} and, as shown in Ref.~\cite{NKtop}, 
the differences in the prescription formalism, as well as the treatment of 
the kinematics, are much bigger 
than higher order terms at NNNLO and beyond. Hence the results of 
Ref.~\cite{BCMN} are quite different from Refs.~\cite{NKtop, KLMV}, 
both theoretically and numerically.

The approach in Refs.~\cite{NKtop,KLMV} was later improved by adding NNNLL 
terms at NNLO \cite{NKRVtop}. 
Although the complete NNNLL terms require calculation of the two-loop 
soft anomalous dimension matrix, it was clearly demonstrated \cite{NKRVtop} 
that the contribution of this matrix at two loops is expected to be negligible.
Thus it is possible to obtain an effective NNNLL calculation by including 
all other terms.  The $\zeta$ terms arising from the inversion to momentum 
space (including some $\zeta$ virtual terms) are dominant, as shown by 
expressing the partonic cross sections in terms of scaling functions that 
depend on the variable $\eta=s/(4m^2)-1$ and comparing the 1PI and PIM scaling 
functions over a large range of $\eta$. Since a complete NNLO 
calculation should be independent of the kinematics, the difference between 
the 1PI and PIM results as a function of $\eta$ is an indication of the 
unknown terms. Away from threshold, hard gluon terms contribute. Since their 
form is also kinematics dependent it is inevitable that, as one moves away
from the threshold region, the 1PI and PIM results 
diverge. However, near threshold the soft gluons dominate and thus 
a complete calculation of the soft terms should produce agreement between 
the 1PI and PIM scaling functions.  At NNLL, the 1PI and PIM functions diverge 
already at threshold \cite{KLMV}, indicating that the NNNLL terms are 
non-negligible.  However, when the NNNLL terms were added \cite{NKRVtop}, this 
discrepancy disappeared, 
as one would expect when all NNLO soft terms are included.  Thus
the contribution of the unknown two-loop soft-anomalous-dimension terms 
that were left out is negligible and we obtain an excellent approximation 
to the complete NNNLL terms and an effective NNNLL calculation, 
denoted NNLO-NNNLL+$\zeta$ in Ref.~\cite{NKRVtop}.

Recently the two-loop soft anomalous dimension for massless quark scattering 
was completed \cite{ADS}. (Work is in progress for heavy quark 
production \cite{NKPS}.)  It was shown \cite{ADS} that the two-loop soft 
anomalous dimension is simply the one-loop result multiplied by half the 
two-loop quantity $K$ \cite{JKLT}.  Assuming that this relation also holds 
for heavy quarks, consistent with 
the two-loop results in Ref.~\cite{SMPU}, the contribution
of this additional two-loop term to the total cross section is less than 1 
per mille at both the 
Tevatron and LHC energies.  It is thus insignificant relative to the 
size of other terms and sources of uncertainty, as expected \cite{NKRVtop}, 
verifying the robustness of the calculation in Ref.~\cite{NKRVtop}.

A very recent paper \cite{SMPU} uses the general approach of Ref.~\cite{BCMN},  
extending the results of Ref.~\cite{BCMN} by adding 
the NNLL terms in the resummed expression. 
A NNLO expansion in powers of $\ln\beta$, where $\beta=\sqrt{1-4m^2/s}$,
is also presented in Ref.~\cite{SMPU}. 
An additional two-loop term is also included, a rough analog of the two-loop 
soft anomalous dimension term in the formalism of Refs.~\cite{NKGS,ADS,NKPS}. 
This two-loop term is again given by 
the one-loop result multiplied by the two-loop quantity $K$ \cite{JKLT}, 
analogous to the result of Ref.~\cite{ADS}. 
We have investigated the contribution of this two-loop term within the approach
of Ref.~\cite{SMPU} and find it to be numerically negligible, on the order of 
a few per mille, consistent 
with the study mentioned previously and again verifying that this two-loop 
contribution is numerically insignificant. Since \cite{SMPU} uses the 
formalism and approximation of Ref.~\cite{BCMN}, their results differ from 
ours for the reasons explained earlier. 
We use exact kinematics in a double differential cross section and define 
partonic threshold through the quantities $s_4$ or $z$, depending on the 
kinematics choice, while Ref.~\cite{SMPU} defines threshold at the total cross 
section level only in terms of $\beta$.

We also note that the authors of Ref.~\cite{SMPU} use the exact scale 
dependence of the NNLO cross section, finding a smaller scale dependence than 
in Refs.~\cite{BCMN,CFMNR}.
The exact NNLO scale dependence was first calculated in Ref.~\cite{KLMV}
and shown to crucially depend on the kinematics 
choice (1PI or PIM).  Hence the scale variation in Refs.~\cite{BCMN,SMPU} 
cannot be directly compared with that of Ref.~\cite{KLMV} or the present paper. 

We also find it more consistent to use the same level of accuracy for the 
scale-dependent terms as for the other terms in the calculation, also chosen 
for the final results in Refs.~\cite{KLMV,NKRVtop}.
We obtain a smaller scale dependence at the Tevatron than Ref.~\cite{SMPU} 
but a larger one at the LHC.
In our approach, the kinematics ambiguity is bigger than the 
scale variation at the Tevatron but smaller at the LHC.
The results in Refs.~\cite{BCMN,CFMNR,SMPU} do not have a kinematics 
uncertainty because their approximation is insensitive to the kinematics choice.
Therefore the scale dependence in those approaches, in particular the small 
scale uncertainty at the LHC of Ref.~\cite{SMPU}, is not necessarily 
indicative of the true theoretical uncertainty in the cross section.

In Ref.~\cite{SMPU} subleading Coulomb terms, calculated in \cite{HQFF}, 
were included in the numerical results. In Ref.~\cite{NKRVtop} some, but 
not all, of these terms were included.  We find that these additional 
contributions are completely negligible at the Tevatron, and make a very 
small contribution, included in our new results, at the LHC.

In this paper, we present detailed results for the top quark cross section
at the Tevatron and the LHC. We primarily use the theoretical approach 
of Ref.~\cite {NKRVtop} with some changes and refinements, described in
the text, along with the newest available parton distribution functions (PDFs). 
We also provide a detailed study of theoretical uncertainties 
including kinematics, scale, and PDF uncertainties, as well as a
discussion of other sources.
In Section 2 we provide results for the $t{\bar t}$ production cross section 
in $p{\bar p}$ collisions at the Tevatron at $\sqrt{S}=1.96$ TeV. In 
Section 3 we give the $t{\bar t}$ production cross section in $pp$ 
collisions at the LHC at $\sqrt{S}=14$ TeV as well as a prediction for 10 TeV.

\mysection{The top quark cross section at the Tevatron}

We begin with $t{\bar t}$ production at the Tevatron at $\sqrt{S}=1.96$ TeV. 
The leading-order partonic processes are  $q{\bar q} \rightarrow t{\bar t}$ 
and $gg \rightarrow  t{\bar t}$. In $p \overline p$ collisions at the 
Tevatron, the $q{\bar q}$ channel is dominant.  In addition to corrections 
to the LO processes, at NLO there are small contributions from two additional 
processes, $qg$ and ${\bar q}g$.

We first calculate the NLO cross section \cite{NLO}, including all channels. 
We then add the NNLO soft-gluon corrections in the  $q{\bar q}$ 
and $gg$ channels to the NLO result. We calculate the soft-gluon corrections 
in both 1PI and PIM kinematics. We find that the behavior of the $q{\bar q} 
\rightarrow t{\bar t}$ and $gg \rightarrow  t{\bar t}$ contributions is quite 
different.  The $q{\bar q}$ channel, with only one diagram at LO, is well 
behaved in both kinematics. 
The $gg$ channel is, however, better treated in 1PI kinematics because the 
three LO $gg \rightarrow t \overline t$ diagrams favor 1PI kinematics. 
In addition, in PIM kinematics the one-loop expansion of the resummed $gg$ 
cross section is very different from the exact NLO result while the 
1PI expansion of the $gg$ contribution is an excellent approximation to the 
exact NLO result at both the Tevatron and the LHC.
Therefore, for our best prediction of the NNLO corrections, we take the 
average of the 1PI and PIM $q{\bar q}$ soft-gluon corrections and add it to 
the 1PI $gg$ soft-gluon result. This is a refinement of our previous method 
\cite{NKRVtop} where we averaged the 1PI and PIM results from both channels. 
Our new approach results in a slightly larger total cross section with a 
somewhat reduced kinematics uncertainty at the Tevatron than in 
Ref.~\cite{NKRVtop}.  As discussed in the Introduction, the effect of 
the two-loop soft anomalous dimension matrix and 
further subleading Coulomb terms is negligible with no change on the results 
presented here.  

We present two tables with NLO and approximate NNLO cross sections using 
the MRST 2006 NNLO \cite{MRST2006} and CTEQ6.6M \cite{CTEQ66} parton densities. 
While the NNLO approximate cross section is of the same logarithmic accuracy 
as the NNLO-NNNLL+$\zeta$ cross section of Ref.~\cite{NKRVtop}, they differ 
slightly because of the changes and refinements discussed above. All the
results are in the $\overline{\rm MS}$ scheme.

\begin{table}[htb]
\begin{center}
\begin{tabular}{|c|c|c|} \hline
\multicolumn{3}{|c|}{MRST 2006 NNLO} \\ \hline 
Mass & $\sigma$(NLO $\pm$ scale & $\sigma$(NNLO approx $\pm$ kinematics  \\
(GeV) & $\pm$ PDF) (pb) & $\pm$ scale $\pm$ PDF) (pb) \\ \hline
165 & $9.23\;^{+0.59}_{-1.09}\;{}^{+0.27}_{-0.23}$ & 
$9.80 \; \pm 0.38 \;^{+0.04}_{-0.34}\;{}^{+0.29}_{-0.25}$  \\ \hline
166 & $8.93\;^{+0.57}_{-1.06}\;{}^{+0.27}_{-0.22}$ &   
$9.48 \; \pm 0.37 \;^{+0.04}_{-0.33}\;{}^{+0.29}_{-0.24}$  \\ \hline
167 & $8.65\;^{+0.55}_{-1.02}\;{}^{+0.26}_{-0.21}$ & 
$9.17 \; \pm 0.36 \;^{+0.04}_{-0.32}\;{}^{+0.28}_{-0.22}$  \\ \hline
168 & $8.37\;^{+0.53}_{-0.99}\;{}^{+0.25}_{-0.20}$ &
$8.88 \; \pm 0.35 \;^{+0.04}_{-0.31}\;{}^{+0.27}_{-0.21}$  \\ \hline
169 & $8.11\;^{+0.51}_{-0.96}\;{}^{+0.24}_{-0.19}$ &
$8.60 \; \pm 0.34 \;^{+0.03}_{-0.30}\;{}^{+0.26}_{-0.20}$  \\ \hline
170 & $7.85\;^{+0.50}_{-0.93}\;{}^{+0.23}_{-0.19}$ &
$8.32 \; \pm 0.33 \;^{+0.03}_{-0.29}\;{}^{+0.25}_{-0.20}$  \\ \hline
171 & $7.60\;^{+0.48}_{-0.90}\;{}^{+0.23}_{-0.18}$ &
$8.06 \; \pm 0.32 \;^{+0.03}_{-0.28}\;{}^{+0.24}_{-0.19}$  \\ \hline
172 & $7.36\;^{+0.46}_{-0.87}\;{}^{+0.22}_{-0.18}$ &
$7.80 \; \pm 0.31 \;^{+0.03}_{-0.27}\;{}^{+0.23}_{-0.19}$  \\ \hline
173 & $7.13\;^{+0.45}_{-0.84}\;{}^{+0.21}_{-0.17}$ &
$7.56 \; \pm 0.30 \;^{+0.02}_{-0.26}\;{}^{+0.22}_{-0.18}$  \\ \hline
174 & $6.91\;^{+0.44}_{-0.82}\;{}^{+0.20}_{-0.17}$ &
$7.32 \; \pm 0.29 \;^{+0.02}_{-0.26}\;{}^{+0.21}_{-0.18}$  \\ \hline
175 & $6.70\;^{+0.42}_{-0.79}\;{}^{+0.19}_{-0.16}$ &
$7.09 \; \pm 0.28 \;^{+0.02}_{-0.25}\;{}^{+0.20}_{-0.17}$  \\ \hline
176 & $6.49\;^{+0.41}_{-0.77}\;{}^{+0.19}_{-0.15}$ &
$6.87 \; \pm 0.27 \;^{+0.02}_{-0.24}\;{}^{+0.20}_{-0.16}$  \\ \hline
177 & $6.29\;^{+0.39}_{-0.74}\;{}^{+0.18}_{-0.15}$ &
$6.66 \; \pm 0.26 \;^{+0.02}_{-0.23}\;{}^{+0.19}_{-0.16}$  \\ \hline
178 & $6.10\;^{+0.38}_{-0.72}\;{}^{+0.18}_{-0.14}$ &
$6.46 \; \pm 0.26 \;^{+0.02}_{-0.23}\;{}^{+0.19}_{-0.15}$  \\ \hline
179 & $5.91\;^{+0.37}_{-0.70}\;{}^{+0.17}_{-0.14}$ &
$6.26 \; \pm 0.25 \;^{+0.02}_{-0.22}\;{}^{+0.18}_{-0.15}$  \\ \hline
180 & $5.73\;^{+0.36}_{-0.68}\;{}^{+0.17}_{-0.13}$ &
$6.07 \; \pm 0.24 \;^{+0.01}_{-0.21}\;{}^{+0.18}_{-0.14}$ \\ \hline
\end{tabular}
\caption[]{The $t \overline t$ production cross section in
$p \overline p$ collisions at the Tevatron with $\sqrt{S}=1.96$ TeV
using the MRST 2006 NNLO PDFs.
The exact NLO results are shown with the scale and PDF uncertainties while 
the approximate NNLO results include kinematics, scale, and PDF uncertainties.}
\label{table1}
\end{center}
\end{table}

Table~\ref{table1} provides the $p \overline p \rightarrow
t \overline t$ cross section for 
$165 < m < 180$ GeV, in 1 GeV increments calculated with the MRST 2006 NNLO
PDFs. We give both the exact NLO and the approximate NNLO cross sections. 
The central values are calculated with the factorization scale, $\mu_F$,
and the renormalization scale, $\mu_R$, set equal to the top quark mass,
$\mu_F = \mu_R = \mu = m$, using the central MRST 2006 NNLO PDF. 

In addition to the central value of the NLO cross section, we also provide 
uncertainties due to the scale and PDF variations. The scale is varied over 
$m/2 < \mu < 2m$.  Varying the scale by a factor of two around $\mu = m$ is a 
standard but arbitrary way to estimate uncertainties from higher-order 
terms.
The $+$ ($-$) indicates the difference between the 
calculation with $\mu=m/2$ ($\mu=2m$) and the central value with $\mu=m$. 
The PDF uncertainty, calculated using the 30 different MRST 2006 NNLO eigensets,
is relatively large, reflecting the uncertainty in the large $x$ region of
the PDFs.

The central value of the NNLO approximate cross section is followed by the 
kinematics uncertainty, the scale variation and the PDF uncertainty.
The NNLO scale and PDF uncertainties are obtained the same way as for the NLO
cross section.  The kinematics uncertainty resides in the treatment of the
$q \overline q$ channel because the $gg$ channel is only calculated in 1PI
kinematics.  The central value is obtained from the average of the 1PI and PIM
$q \overline q$ calculations.  The $+$ kinematics uncertainty is the found by
taking the 1PI $q \overline q$ result alone while the $-$ results uses the PIM
$q \overline q$ calculation.  The kinematics uncertainty is symmetric becuase
the central $q \overline q$ contribution is the average of the 1PI and PIM
results.

At NLO, the scale variation is significant. When the NNLO corrections are 
added, the scale dependence on the scale decreases dramatically.  However, 
the kinematics uncertainty is larger than the scale variation.  The PDF 
uncertainty is also significant at NLO and NNLO, of the same order as the 
kinematics dependence at NNLO.  

\begin{table}[htb]
\begin{center}
\begin{tabular}{|c|c|c|} \hline
\multicolumn{3}{|c|}{CTEQ6.6M} \\ \hline 
Mass & $\sigma$(NLO $\pm$ scale & $\sigma$(NNLO approx $\pm$ kinematics  \\
(GeV) & $\pm$ PDF) (pb) & $\pm$ scale $\pm$ PDF) (pb) \\ \hline
165 & $8.74\;^{+0.46}_{-0.96}\;{}^{+0.58}_{-0.45}$ &  
$9.23 \; \pm 0.37 \;^{-0.03}_{-0.25}\;{}^{+0.61}_{-0.48}$ \\ \hline
166 & $8.47\;^{+0.44}_{-0.93}\;{}^{+0.56}_{-0.43}$ &  
$8.93 \; \pm 0.36 \;^{-0.03}_{-0.24}\;{}^{+0.59}_{-0.45}$    \\ \hline
167 & $8.20\;^{+0.43}_{-0.90}\;{}^{+0.53}_{-0.42}$ &  
$8.65 \; \pm 0.35 \;^{-0.03}_{-0.23}\;{}^{+0.56}_{-0.44}$    \\ \hline
168 & $7.94\;^{+0.42}_{-0.87}\;{}^{+0.52}_{-0.41}$ &  
$8.38 \; \pm 0.34 \;^{-0.03}_{-0.23}\;{}^{+0.55}_{-0.43}$    \\ \hline
169 & $7.70\;^{+0.40}_{-0.84}\;{}^{+0.50}_{-0.39}$ &  
$8.12 \; \pm 0.33 \;^{-0.03}_{-0.22}\;{}^{+0.53}_{-0.41}$    \\ \hline
170 & $7.46\;^{+0.39}_{-0.82}\;{}^{+0.48}_{-0.38}$ &  
$7.87 \; \pm 0.32 \;^{-0.03}_{-0.21}\;{}^{+0.51}_{-0.40}$    \\ \hline
171 & $7.23\;^{+0.38}_{-0.79}\;{}^{+0.47}_{-0.36}$ &  
$7.62 \; \pm 0.31 \;^{-0.03}_{-0.21}\;{}^{+0.50}_{-0.38}$    \\ \hline
172 & $7.01\;^{+0.37}_{-0.77}\;{}^{+0.45}_{-0.35}$ &  
$7.39 \; \pm 0.30 \;^{-0.03}_{-0.20}\;{}^{+0.48}_{-0.37}$    \\ \hline
173 & $6.79\;^{+0.35}_{-0.74}\;{}^{+0.43}_{-0.34}$ &  
$7.16 \; \pm 0.29 \;^{-0.03}_{-0.19}\;{}^{+0.45}_{-0.36}$    \\ \hline
174 & $6.58\;^{+0.34}_{-0.72}\;{}^{+0.42}_{-0.33}$ &  
$6.94 \; \pm 0.28 \;^{-0.03}_{-0.19}\;{}^{+0.44}_{-0.35}$    \\ \hline
175 & $6.38\;^{+0.33}_{-0.70}\;{}^{+0.41}_{-0.31}$ &  
$6.73 \; \pm 0.27 \;^{-0.03}_{-0.18}\;{}^{+0.43}_{-0.33}$    \\ \hline
176 & $6.19\;^{+0.32}_{-0.68}\;{}^{+0.39}_{-0.30}$ &  
$6.53 \; \pm 0.27 \;^{-0.03}_{-0.18}\;{}^{+0.41}_{-0.32}$    \\ \hline
177 & $6.00\;^{+0.31}_{-0.66}\;{}^{+0.38}_{-0.29}$ &  
$6.33 \; \pm 0.26 \;^{-0.03}_{-0.17}\;{}^{+0.40}_{-0.31}$    \\ \hline
178 & $5.82\;^{+0.30}_{-0.64}\;{}^{+0.37}_{-0.28}$ &  
$6.14 \; \pm 0.25 \;^{-0.03}_{-0.17}\;{}^{+0.39}_{-0.30}$    \\ \hline
179 & $5.65\;^{+0.29}_{-0.62}\;{}^{+0.35}_{-0.27}$ &  
$5.95 \; \pm 0.24 \;^{-0.03}_{-0.16}\;{}^{+0.37}_{-0.28}$    \\ \hline
180 & $5.48\;^{+0.28}_{-0.60}\;{}^{+0.34}_{-0.26}$ &  
$5.77 \; \pm 0.24 \;^{-0.03}_{-0.16}\;{}^{+0.36}_{-0.27}$    \\ \hline
\end{tabular}
\caption[]{The $t \overline t$ production cross section in
$p \overline p$ collisions at the Tevatron with $\sqrt{S}=1.96$ TeV
using the CTEQ6.6M PDFs.
The exact NLO results are shown with the scale and PDF uncertainties while
the approximate NNLO results include kinematics, scale, and PDF uncertainties.}
\label{table2}
\end{center}
\end{table}

Table~\ref{table2} provides the $t \overline t$ cross section for $165 < m < 
180$ GeV, in 1 GeV increments, using the CTEQ6.6M NLO PDFs.
As in Table 1,  we list both the exact NLO and our
approximate NNLO cross sections together with all uncertainties.
The central values, again shown with $\mu_F = \mu_R = \mu = m$, employ
the central CTEQ6.6M PDFs.

The scale variation is calculated as described above for the MRST 2006 NNLO
PDFs.  In the case of the NNLO approximate cross section, the NNLO results at
both ends of the scale range, $\mu = m/2$ and $2m$, are lower than with 
$\mu=m$, indicated by the double minus signs on the scale uncertainty.
The PDF uncertainty is calculated using the 44 different CTEQ6.6M eigensets.

The two sets of results are quite different.  The cross sections calculated with
CTEQ6.6M are smaller than those with the central MRST 2006 NNLO set but have
larger PDF uncertainties. Indeed,
the CTEQ6.6M NNLO approximate cross sections are quite similar to the NLO
cross section calculated with the MRST 2006 NNLO PDFs.  Part of the difference
can be attributed to the fact that the MRST 2006 NNLO sets are of the same order
as the NNLO approximate calculation while the CTEQ6.6M sets are NLO.
Furthermore, the CTEQ6.6M large-$x$ gluon distribution is smaller, reducing the
relative $gg$ contribution.

The best way to combine the uncertainties is not obvious. The most 
conservative approach would be to add them linearly. However the kinematics 
and scale uncertainties both reflect the neglect of uknown terms.  Thus
a linear combination of the uncertainties likely provides an overestimate of
the overall uncertainty and we instead prefer to add them in quadrature. 

We present the NNLO approximate cross section for the current most likely 
value of the top quark mass, $m=172$ GeV for both sets of PDFs.
We first give the central result from Tables~\ref{table1} and \ref{table2}
with all the uncertainties shown separately and then add the 
uncertainties in quadrature for our final result.
Using the MRST 2006 NNLO PDFs, we have 
\beq
\sigma^{\rm NNLOapprox}_{p{\bar p} \rightarrow t \bar t}(1.96\, {\rm TeV},
m=172 \, 
{\rm GeV}, {\rm MRST})=7.80 \; \pm 0.31 \;^{+0.03}_{-0.27}\;{}^{+0.23}_{-0.19}
\; {\rm pb}=7.80\; {}^{+0.39}_{-0.45} \; {\rm pb} \, ,
\eeq
while with CTEQ6.6M we find
\beq
\sigma^{\rm NNLOapprox}_{p{\bar p} \rightarrow t \bar t}(1.96\, {\rm TeV},
m=172 \, 
{\rm GeV}, {\rm CTEQ})=7.39 \; \pm 0.30 \;^{-0.03}_{-0.20}\;{}^{+0.48}_{-0.37}
\; {\rm pb}=7.39\; {}^{+0.57}_{-0.52} \; {\rm pb} \, .
\eeq

We have not included further theoretical ambiguities arising from the choice 
of equivalent analytical expressions near threshold or from damping factors 
\cite{KLMV} as well as from the virtual $\zeta$ terms discussed in 
Ref.~\cite{NKRVtop}.  Such ambiguities are partly accounted for in the 
kinematics and scale uncertainties shown here.

\begin{figure}
\begin{center}
\hspace{-5mm}
\includegraphics[width=8cm]{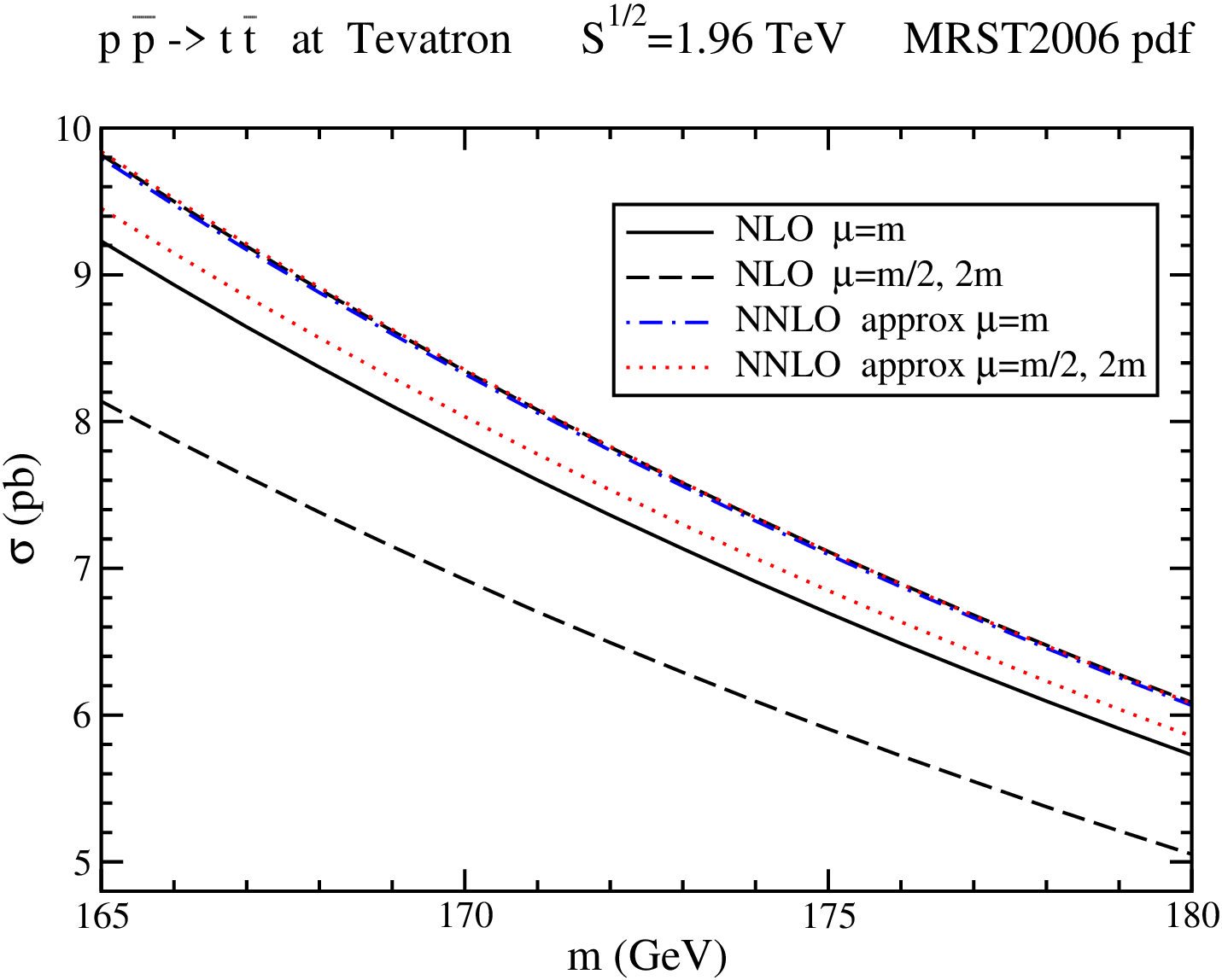}
\hspace{5mm}
\includegraphics[width=8cm]{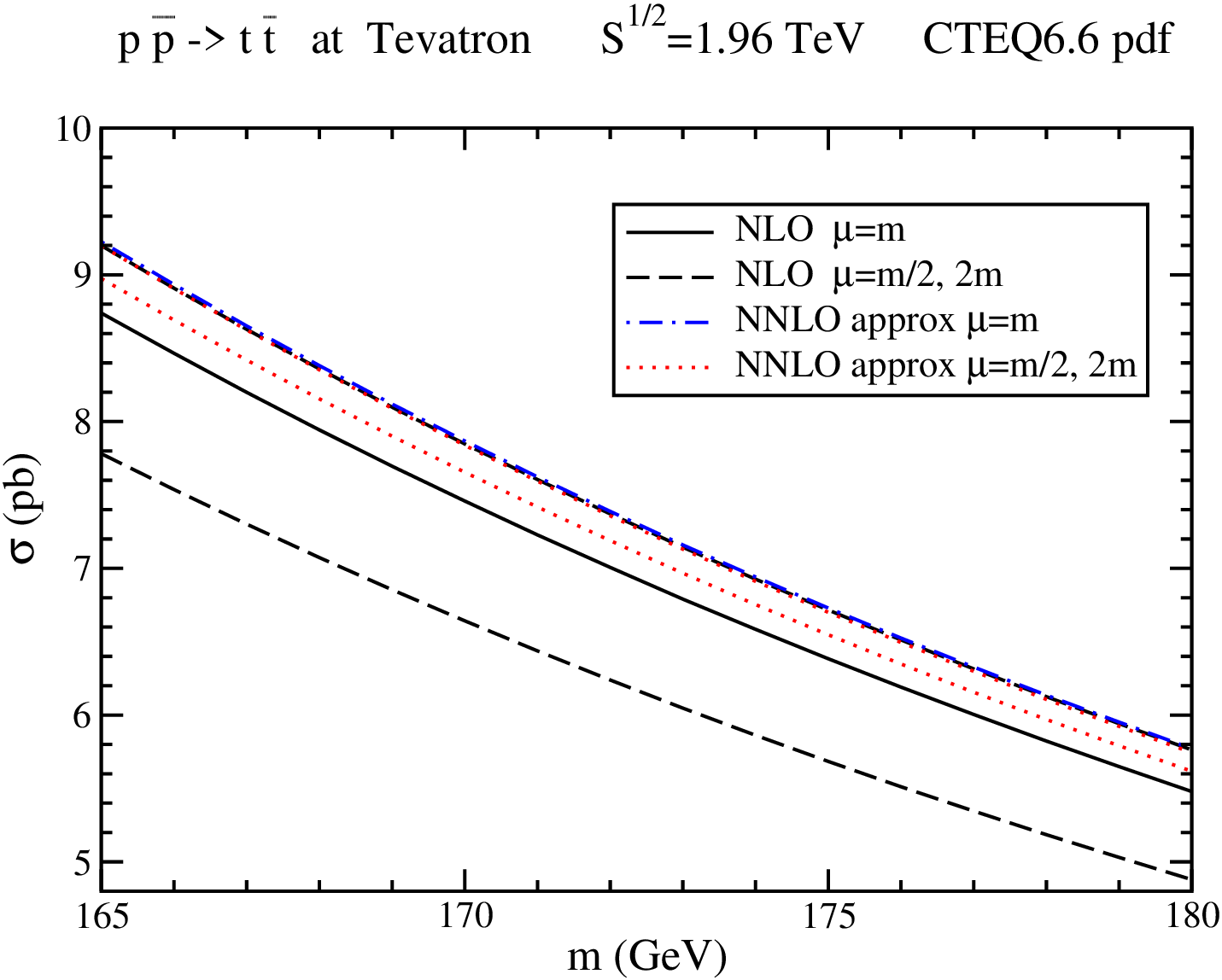}
\caption{The exact NLO and approximate NNLO cross sections in $p \overline p$
collisions at 1.96 TeV
using the MRST 2006 NNLO (left) and CTEQ6.6M (right) PDFs.}
\label{NNLOtevplot}
\end{center}
\end{figure}

Figure~\ref{NNLOtevplot} shows the exact NLO and approximate NNLO top quark 
cross sections as a function of top quark mass at the Tevatron using the MRST 
2006 NNLO (left) and CTEQ6.6M (right) PDFs.   
Three curves are given for each order: a central value with $\mu=m$ and the
extremes of the calculated scale dependence with $\mu=m/2$ and $2m$.  The 
region between the upper and lower scales represents the scale variation at 
each order. We see that the NNLO scale dependence is 
much diminished relative to NLO. In fact the NNLO curves with $\mu=m/2$ 
and $\mu=m$ are on top of each other as well as on top of the NLO curve with 
$\mu=m/2$.  The kinematics and PDF uncertainties are not represented in the 
plots.

\begin{figure}
\begin{center}
\hspace{-5mm}
\includegraphics[width=8cm]{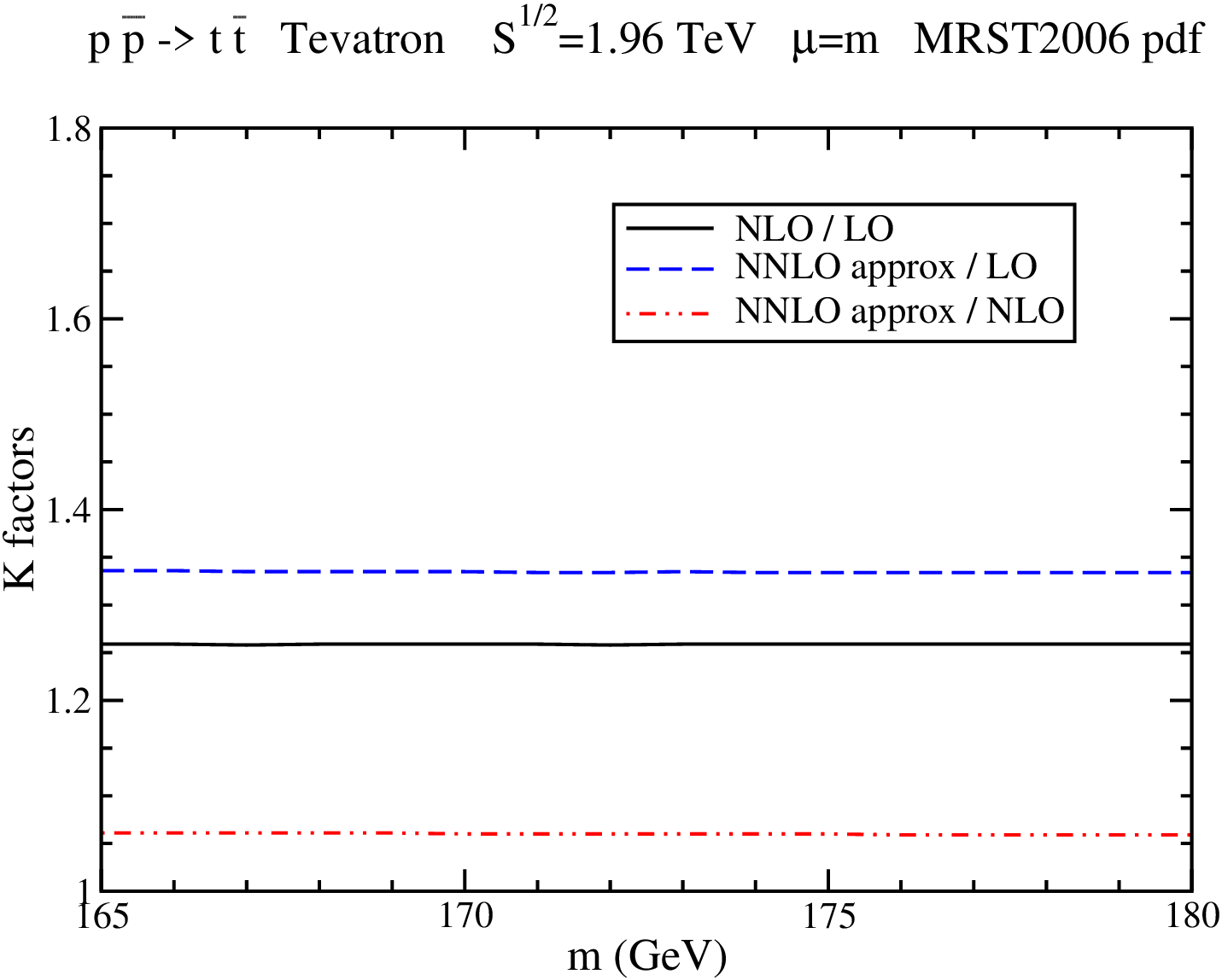}
\hspace{5mm}
\includegraphics[width=8cm]{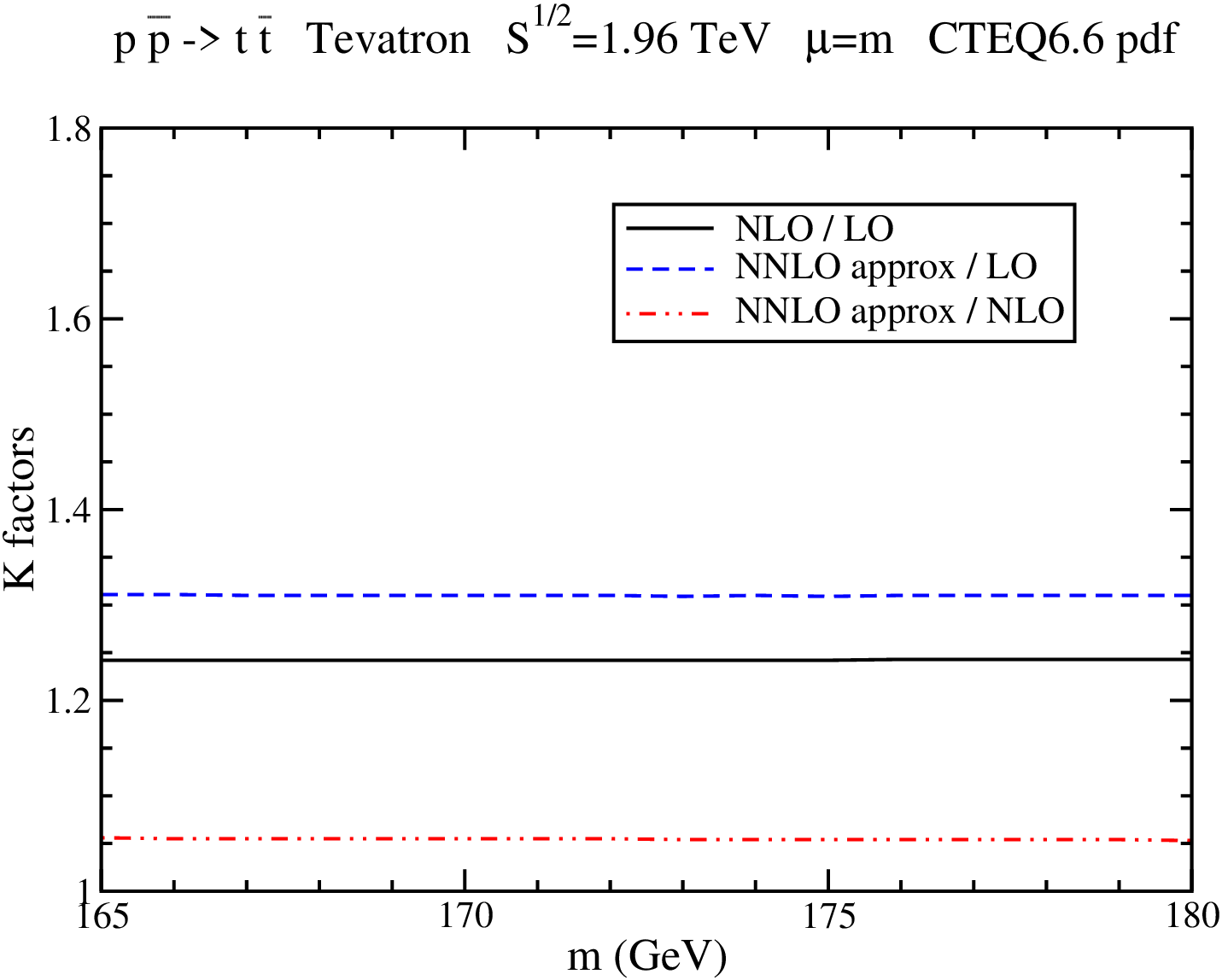}
\caption{The $K$ factors at the Tevatron using the MRST 2006 NNLO (left) 
and the CTEQ6.6M (right) PDFs.}
\label{Ktevplot}
\end{center}
\end{figure}

In Fig.~\ref{Ktevplot}, we present the $K$ factors at the Tevatron using the 
MRST 2006 NNLO (left) and CTEQ6.6M (right) PDFs with our central value, $\mu=m$.
The $K$ factors are virtually independent of the PDF although the CTEQ6.6M 
$K$ factors appear slightly smaller. They are also independent of the top quark
mass.  The ratios of the approximate NNLO cross sections to the exact LO and
NLO cross sections are both given.  The NNLO corrections enhance the NLO
$t \overline t$ cross section by $\sim 6$\% at $\mu=m$.

\mysection{The top quark cross section at the LHC}

We now turn to $t{\bar t}$ production in $pp$ collisions at the LHC. 
While our results are primarily shown
for $\sqrt{S}=14$ TeV, we also provide predictions for the top quark cross
section at the LHC start-up energy of 10 TeV.  We note that at the LHC, 
the $gg$ channel is dominant.

\begin{table}[htb]
\begin{center}
\begin{tabular}{|c|c|c|} \hline
\multicolumn{3}{|c|}{MRST 2006 NNLO} \\ \hline
Mass & $\sigma$(NLO $\pm$ scale & $\sigma$(NNLO approx $\pm$ kinematics  \\
(GeV) & $\pm$ PDF) (pb) & $\pm$ scale $\pm$ PDF) (pb) \\ \hline
165 & $1089\;^{+135}_{-129}\;^{+12}_{-14}$ & 
$1173 \; \pm 5 \;^{+95}_{-62}\;{}^{+13}_{-15}$  \\ \hline
166 & $1059\;^{+132}_{-125}\;^{+12}_{-13}$ &
$1141 \; \pm 5 \;^{+93}_{-60}\;{}^{+13}_{-14}$  \\ \hline
167 & $1030\;^{+128}_{-122}\;^{+12}_{-13}$ & 
$1109 \; \pm 5 \;^{+90}_{-58}\;{}^{+13}_{-14}$  \\ \hline
168 & $1003\;^{+124}_{-119}\;^{+12}_{-13}$ & 
$1080 \; \pm 5 \;^{+87}_{-57}\;{}^{+13}_{-14}$  \\ \hline
169 & $ 976\;^{+120}_{-116}\;^{+12}_{-12}$ & 
$1050 \; \pm 5 \;^{+85}_{-55}\;{}^{+13}_{-13}$  \\ \hline
170 & $ 950\;^{+117}_{-113}\;^{+12}_{-12}$ & 
$1022 \; \pm 5 \;^{+83}_{-53}\;{}^{+13}_{-13}$  \\ \hline
171 & $ 924\;^{+114}_{-110}\;^{+12}_{-12}$ & 
$994  \; \pm 5 \;^{+81}_{-52}\;{}^{+13}_{-13}$  \\ \hline
172 & $ 900\;^{+110}_{-107}\;^{+11}_{-12}$ &  
$968 \; \pm  4\;^{+79}_{-50}\;{}^{+12}_{-13}$   \\ \hline
173 & $ 876\;^{+108}_{-104}\;^{+11}_{-11}$ &  
$943 \; \pm  4\;^{+77}_{-49}\;{}^{+12}_{-12}$   \\ \hline
174 & $ 853\;^{+105}_{-101}\;^{+11}_{-11}$ &  
$918 \; \pm  4\;^{+75}_{-48}\;{}^{+12}_{-12}$   \\ \hline
175 & $ 831\;^{+102}_{- 98}\;^{+11}_{-11}$ &  
$894 \; \pm  4\;^{+73}_{-46}\;{}^{+12}_{-12}$   \\ \hline
176 & $ 809\;^{+ 99}_{- 96}\;^{+10}_{-10}$ &  
$871 \; \pm  4\;^{+71}_{-45}\;{}^{+11}_{-11}$   \\ \hline
177 & $ 788\;^{+ 97}_{- 93}\;^{+10}_{-10}$ &  
$848 \; \pm  4\;^{+69}_{-44}\;{}^{+11}_{-11}$   \\ \hline
178 & $ 768\;^{+ 94}_{- 91}\;^{+10}_{-10}$ &  
$826 \; \pm  4\;^{+67}_{-43}\;{}^{+11}_{-11}$   \\ \hline
179 & $ 748\;^{+ 91}_{- 89}\;^{+10}_{-10}$ &  
$805 \; \pm  4\;^{+65}_{-42}\;{}^{+11}_{-11}$   \\ \hline
180 & $ 729\;^{+ 89}_{- 86}\;^{+ 9}_{-10}$ &  
$785 \; \pm  4\;^{+64}_{-40}\;{}^{+10}_{-11}$   \\ \hline
\end{tabular}
\caption[]{The $t \overline t$ production cross section in
$p p$ collisions at the LHC with $\sqrt{S}=14$ TeV using the MRST 2006 
NNLO PDFs.  The exact NLO results are shown with scale and PDF uncertainties 
while the approximate NNLO results include kinematics, scale, and PDF
uncertainties.}
\label{table3}
\end{center}
\end{table}

Table~\ref{table3} provides the top quark cross section for $165<m< 180$ GeV, 
in 1 GeV increments, in $pp$ collisions at the LHC with $\sqrt{S}=14$ TeV 
employing the MRST 2006 NNLO PDFs. 
We list both the exact NLO and approximate NNLO cross sections. 
The central values are given for $\mu_F = \mu_R = \mu = m$ with the central
MRST 2006 NNLO PDFs.

The NLO cross section is shown with the uncertainties due to the scale
variation and the choice of PDF eigenset. As before, we vary the scale
between $\mu = m/2$ and $2m$ and the PDF uncertainty is calculated using the 
30 different MRST 2006 NNLO eigensets.

The central value of the NNLO approximate cross section is accompanied by
uncertainties due to the kinematics, the scale variation and the choice of PDF.
Again, the central value of the approximate NNLO $q \overline q$ contribution
if the average of the 1PI and PIM kinematics choice.  The $+$ ($-$) kinematics 
uncertainty is the difference between the top cross section with the 
$q{\bar q}$ contribution calculated in 1PI (PIM) kinematics.

At NLO the scale variation is large. When the NNLO corrections are
added the dependence on the scale decreases significantly.  Now the kinematics 
uncertainty is much smaller than that due to the scale variation. This is 
because we only use 1PI kinematics for the $gg$ channel, dominant for $pp$
collisions at this energy.  Thus the kinematics uncertainty is only due to
the change in the $q \overline q$ calculation.  The PDF uncertainty is smaller
at the LHC since $x$ is relatively small, in a range where the PDFs are better
known.

\begin{table}[htb]
\begin{center}
\begin{tabular}{|c|c|c|} \hline
\multicolumn{3}{|c|}{CTEQ6.6M} \\ \hline
Mass & $\sigma$(NLO $\pm$ scale & $\sigma$(NNLO approx $\pm$ kinematics  \\
(GeV) & $\pm$ PDF) (pb) & $\pm$ scale $\pm$ PDF) (pb) \\ \hline
165 & $1035\;^{+125}_{-121}\;^{+31}_{-34}$ &  
$1114 \; \pm 5 \;^{+87}_{-55}\;{}^{+33}_{-37}$   \\ \hline
166 & $1007\;^{+121}_{-118}\;^{+31}_{-33}$ &  
$1084 \; \pm 5 \;^{+84}_{-54}\;{}^{+33}_{-35}$    \\ \hline
167 & $ 979\;^{+117}_{-115}\;^{+30}_{-33}$ &  
$1054 \; \pm 5 \;^{+81}_{-52}\;{}^{+32}_{-35}$    \\ \hline
168 & $ 952\;^{+114}_{-112}\;^{+30}_{-32}$ &  
$1025 \; \pm 5 \;^{+79}_{-50}\;{}^{+32}_{-34}$    \\ \hline
169 & $ 927\;^{+110}_{-109}\;^{+29}_{-31}$ &  
$997  \; \pm 4 \;^{+76}_{-49}\;{}^{+31}_{-33}$      \\ \hline
170 & $ 902\;^{+107}_{-106}\;^{+29}_{-30}$ &  
$970 \; \pm  4 \;^{+74}_{-48}\;{}^{+31}_{-32}$      \\ \hline
171 & $ 877\;^{+105}_{-103}\;^{+28}_{-29}$ &  
$944 \; \pm  4 \;^{+72}_{-47}\;{}^{+30}_{-31}$      \\ \hline
172 & $ 854\;^{+102}_{-100}\;^{+27}_{-29}$ &  
$919 \; \pm  4 \;^{+70}_{-45}\;{}^{+29}_{-31}$      \\ \hline
173 & $ 831\;^{+ 99}_{- 97}\;^{+27}_{-29}$ &  
$894 \; \pm  4 \;^{+68}_{-44}\;{}^{+29}_{-31}$      \\ \hline
174 & $ 809\;^{+ 96}_{- 95}\;^{+26}_{-28}$ &  
$870 \; \pm  4 \;^{+66}_{-43}\;{}^{+28}_{-30}$      \\ \hline
175 & $ 788\;^{+ 94}_{- 92}\;^{+26}_{-27}$ &  
$847 \; \pm  4 \;^{+64}_{-42}\;{}^{+28}_{-29}$      \\ \hline
176 & $ 767\;^{+ 91}_{- 90}\;^{+25}_{-27}$ &  
$825 \; \pm  4 \;^{+62}_{-41}\;{}^{+27}_{-29}$      \\ \hline
177 & $ 747\;^{+ 89}_{- 88}\;^{+25}_{-26}$ &  
$803 \; \pm  4 \;^{+60}_{-39}\;{}^{+27}_{-28}$      \\ \hline
178 & $ 727\;^{+ 86}_{- 85}\;^{+25}_{-26}$ &   
$782 \; \pm  4 \;^{+59}_{-38}\;{}^{+27}_{-28}$     \\ \hline
179 & $ 709\;^{+ 84}_{- 83}\;^{+24}_{-25}$ &  
$762 \; \pm  3 \;^{+57}_{-37}\;{}^{+26}_{-27}$      \\ \hline
180 & $ 690\;^{+ 82}_{- 81}\;^{+23}_{-25}$ &   
$742 \; \pm  3 \;^{+55}_{-36}\;{}^{+25}_{-27}$     \\ \hline
\end{tabular}
\caption[]{The $t \overline t$ production cross section in
$p p$ collisions at the LHC with $\sqrt{S}=14$ TeV using the CTEQ6.6M
PDFs.  The exact NLO results are shown with scale and PDF uncertainties 
while the approximate NNLO results include kinematics, scale, and PDF 
uncertainties.}  
\label{table4}
\end{center}
\end{table}

In Table~\ref{table4}, we present the corresponding top cross sections with 
the CTEQ6.6M NLO PDFs.  The CTEQ6.6M results are again smaller than those with 
the MRST 2006 NNLO sets.  While the PDF uncertainty is smaller at the LHC,
the CTEQ6.6 uncertainty is still larger than those with MRST 2006 NNLO.

We now present our predicted NNLO approximate cross section for top production 
in $pp$ collisions at $\sqrt{S} = 14$ TeV with $m = 172$ GeV.  The results are
again given both with the separate uncertainties, as in Tables~\ref{table3} and
\ref{table4}, and with uncertainties added in quadrature. Using the MRST 2006 
NNLO PDFs, we find
\beq
\sigma^{\rm NNLOapprox}_{pp \rightarrow t \bar t}(14\, {\rm TeV},m=172 \, 
{\rm GeV}, {\rm MRST})=968 \; \pm  4\; {}^{+79}_{-50}\;{}^{+12}_{-13}
\; {\rm pb}=968 \; ^{+80}_{-52} \; {\rm pb} \, ,
\eeq
while with the CTEQ6.6M PDFs, we obtain
\beq
\sigma^{\rm NNLOapprox}_{pp \rightarrow t \bar t}(14\, {\rm TeV},m=172 \, 
{\rm GeV}, {\rm CTEQ})=919 \; \pm  4 \; {}^{+70}_{-45}\; {}^{+29}_{-31}  
\; {\rm pb}=919 \; {}^{+76}_{-55} \; {\rm pb} \, .
\eeq

\begin{figure}
\begin{center}
\hspace{-5mm}
\includegraphics[width=8cm]{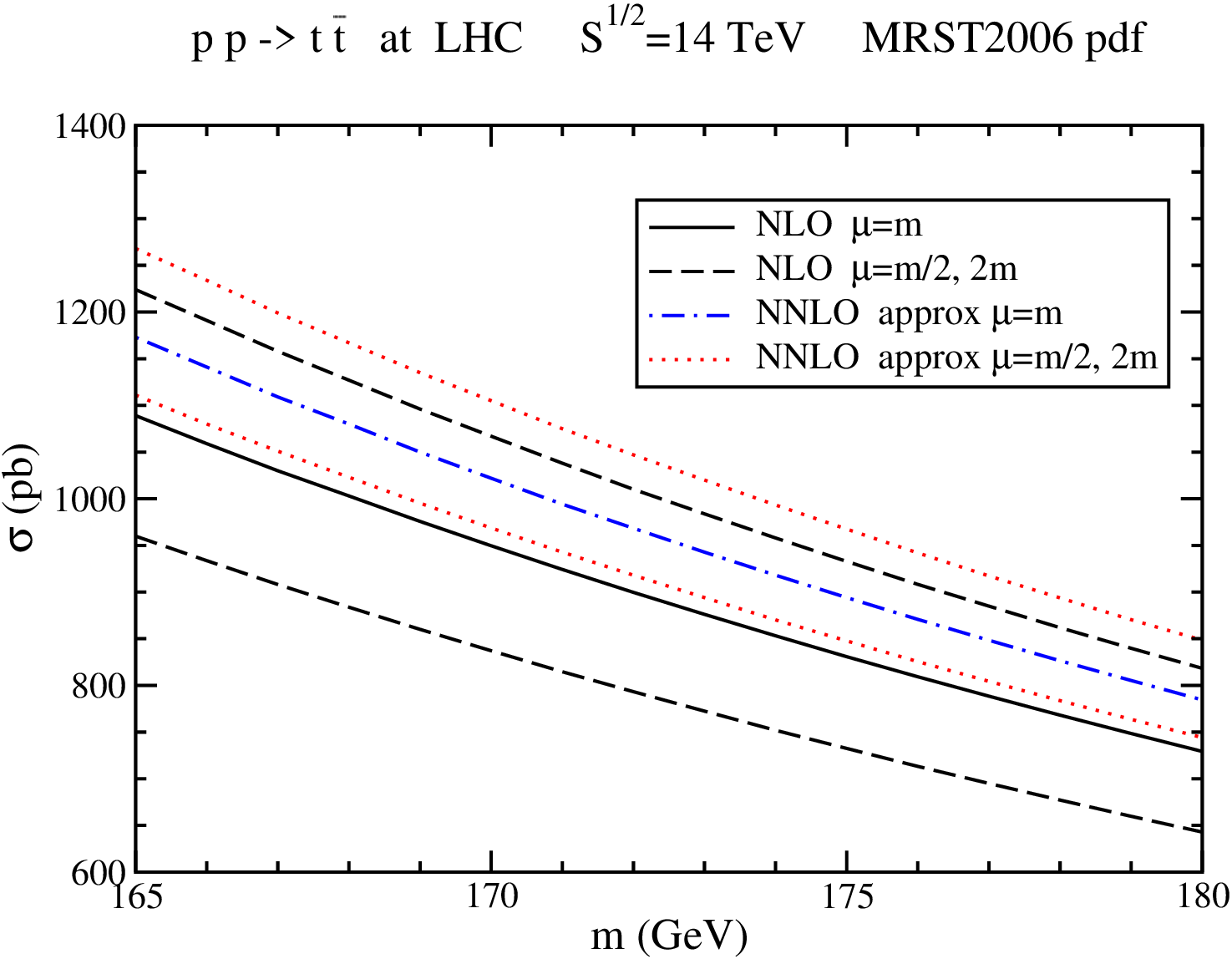}
\hspace{5mm}
\includegraphics[width=8cm]{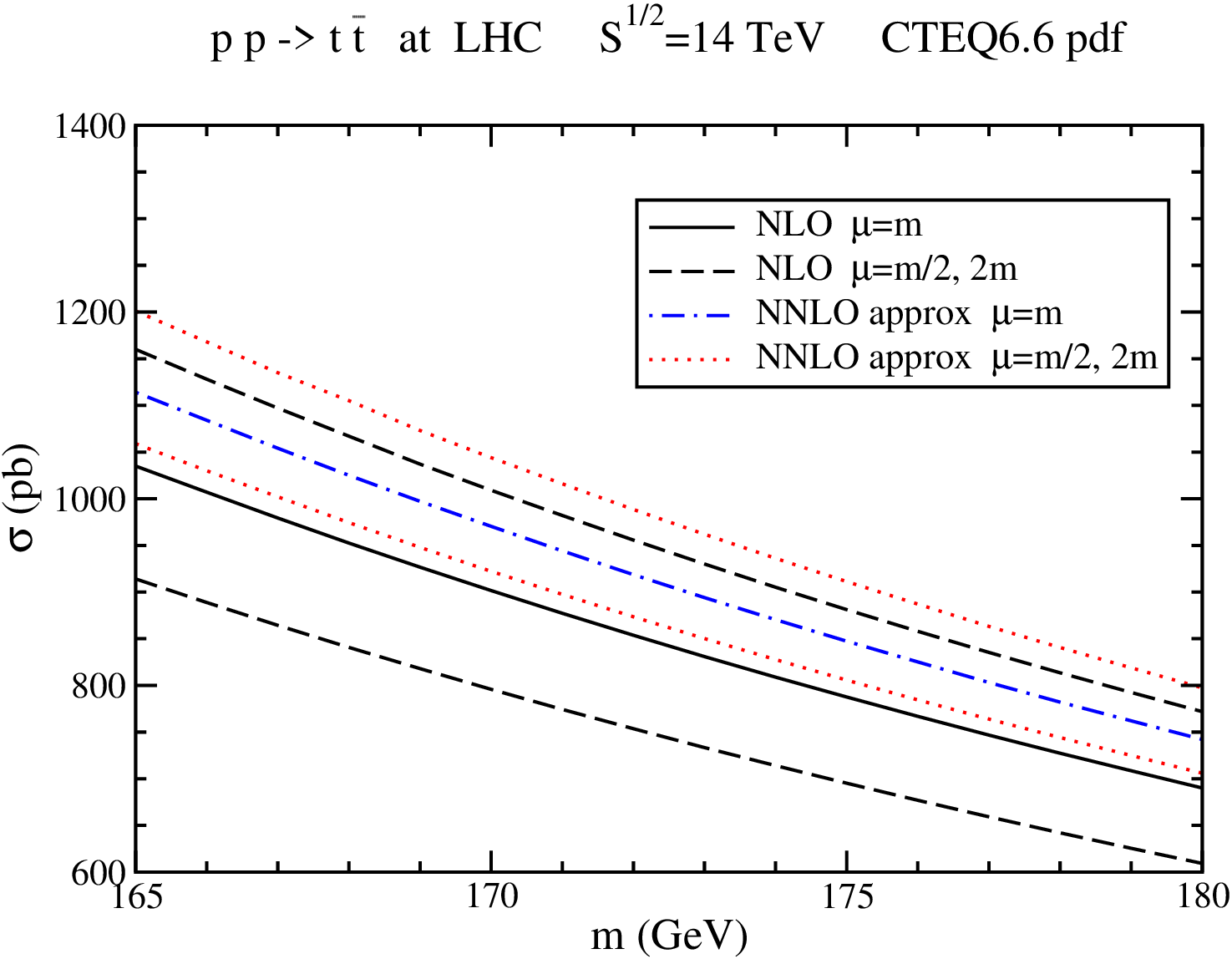}
\caption{The NLO and approximate NNLO top cross sections in 14 TeV $pp$
collisions at the LHC using the MRST 2006 NNLO (left) and the CTEQ6.6M (right) 
PDFs.}
\label{NNLOlhcplot}
\end{center}
\end{figure}

Figure~\ref{NNLOlhcplot} shows exact NLO and approximate NNLO top quark cross 
sections as a function of top quark mass with $\sqrt{S}=14$ TeV using the MRST 
2006 NNLO (left) and CTEQ6.6M (right) PDFs.  At each order we show the central
result with $\mu = m$ as well as the range of the scale uncertainty indicated
by the upper ($\mu = m/2$) and lower ($\mu = 2m$) curves.  The region 
between the upper and lower curves denotes the scale variation.  While the 
NNLO scale dependence is reduced relative to NLO, the reduction is not as
large as at the Tevatron.  We note that the mass dependence at the LHC is
smaller than at the Tevatron since we are further from production threshold
here.

\begin{figure}
\begin{center}
\hspace{-5mm}
\includegraphics[width=8cm]{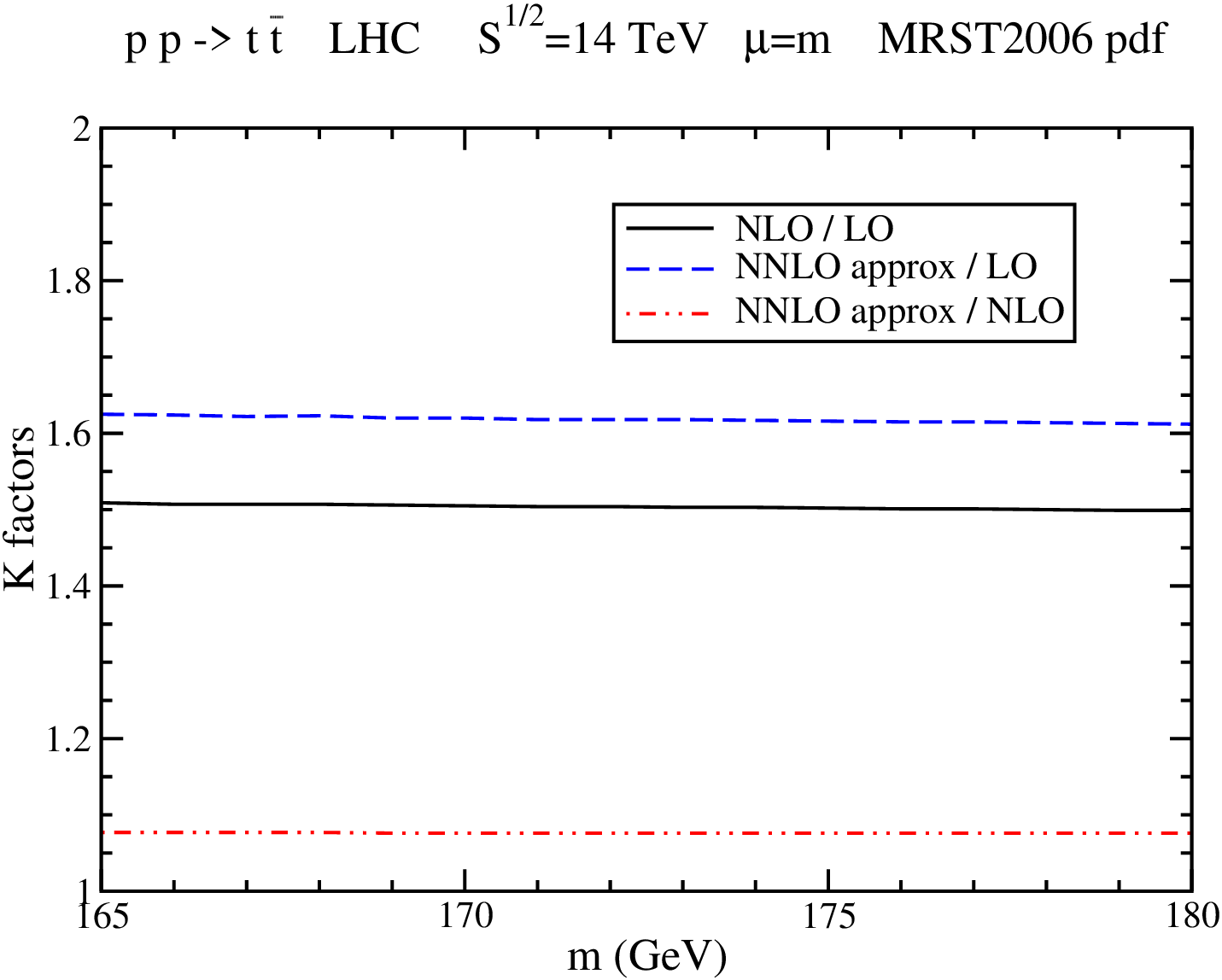}
\hspace{5mm}
\includegraphics[width=8cm]{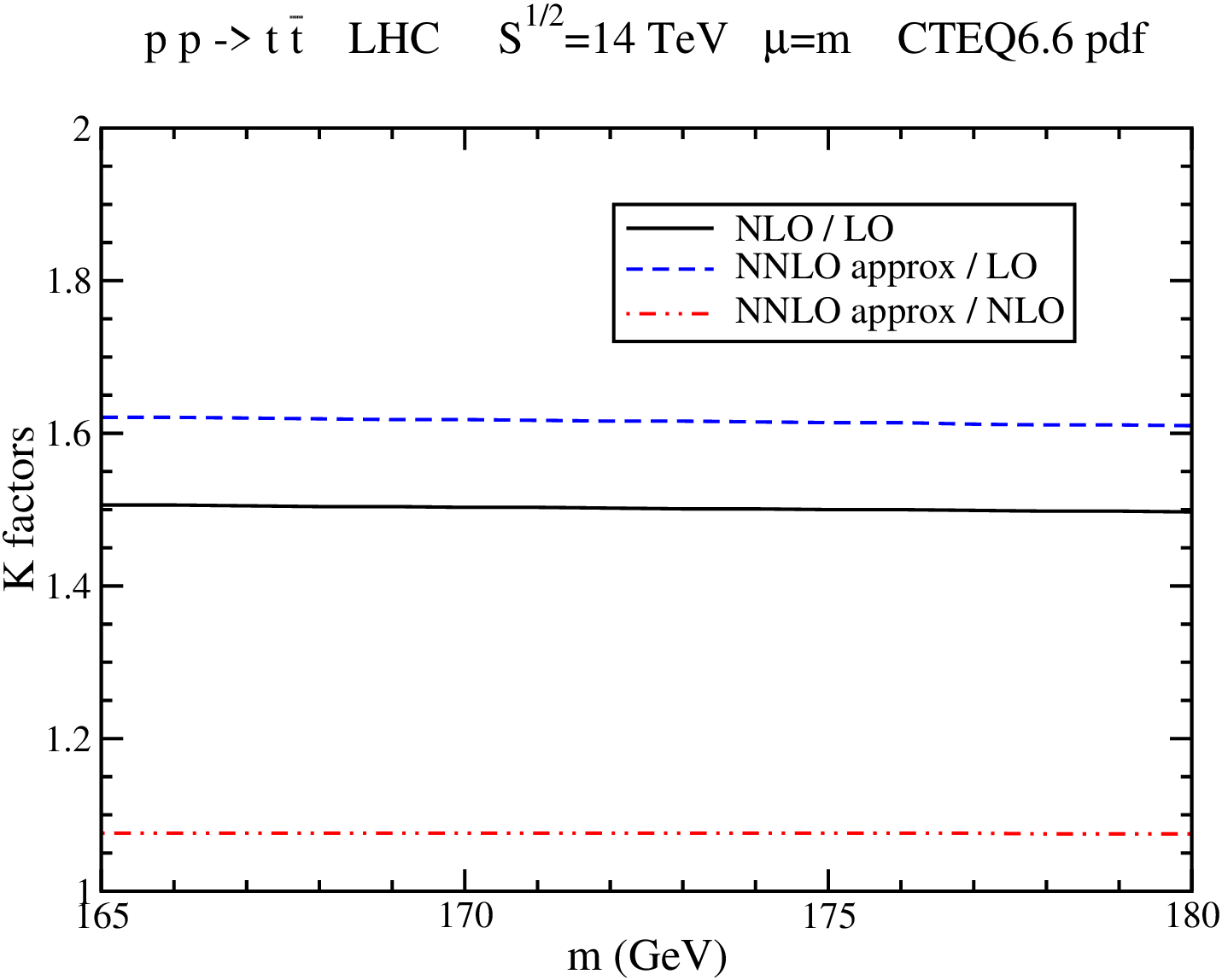}
\caption{The $K$ factors at the LHC using the MRST 2006 NNLO (left) 
and CTEQ6.6M (right) PDFs.}
\label{Klhcplot}
\end{center}
\end{figure}

In Fig.~\ref{Klhcplot}, we show the LHC $K$ factors as a function of mass for
our central ($\mu = m$) cross sections calculated with the MRST 2006 NNLO 
(left) and CTEQ6.6M (right) PDFs.  The LHC $K$ factors are larger than those
shown in Fig.~\ref{Ktevplot}.  They are almost identical for the two sets
and are virtually independent of mass.  The approximate NNLO cross section is
$\sim 8$\% larger than the NLO cross section.

Finally, we provide predictions  for the initial LHC run at $\sqrt{S}=10$ TeV
with $m =172$ GeV. 
Using the MRST 2006 NNLO PDFs, the exact NLO cross section is  
$414\; \pm 52 \; \pm 8$ pb while the NNLO approximate cross section is
\beq
\sigma^{\rm NNLOapprox}_{pp \rightarrow t \bar t}(10\, {\rm TeV},m=172 \, 
{\rm GeV}, {\rm MRST})=446\; \pm 3\; {}^{+32}_{-23} \; {}\pm 9 \; {\rm pb}
=446\; {}^{+33}_{-25} \; {\rm pb} \, .
\eeq
With the CTEQ6.6M PDFs we find that the NLO cross section is 
$385\; {}^{+47}_{-48} \; {}^{+19}_{-18}$ pb and the NNLO approximate cross 
section is 
\beq
\sigma^{\rm NNLOapprox}_{pp \rightarrow t \bar t}(10\, {\rm TeV},m=172 \, 
{\rm GeV}, {\rm CTEQ})=415\; \pm 2\; {}^{+27}_{-21} \; \pm 20 \; {\rm pb} 
=415\; {}^{+34}_{-29} \; {\rm pb} \, .
\eeq

\mysection{Conclusions}

We have studied top quark production at the Tevatron and the LHC.  
Our work is the only calculation that employs full kinematics in the 
double differential cross section beyond NLL using the soft anomalous 
dimension matrix.
We presented detailed results for the exact NLO and approximate 
NNLO $t{\bar t}$ 
cross sections at the Tevatron and the LHC for a wide range of top 
quark masses using the MRST 2006 NNLO  and the CTEQ6.6M PDFs. 
The approximate NNLO corrections include soft-gluon contributions 
which significantly enhance the cross section. We find 
that further two-loop soft gluon contributions are expected to be negligible. 
We also included subleading Coulomb contributions and found them negligible 
at the Tevatron and very small at the LHC.

We provided detailed results for the theoretical uncertainties, including the
kinematics ambiguity, scale variation, and PDF uncertainties. 
We found that the NNLO 
scale uncertainty is drastically reduced relative to NLO at the Tevatron where 
the kinematics uncertainty is larger. The PDF uncertainty is quite significant, 
especially for the CTEQ6.6M PDFs. At the LHC, the kinematics ambiguity is 
small.  The NNLO scale variation is larger despite being significantly 
smaller than at NLO. 
The PDF uncertainty is smaller at the LHC than at the Tevatron. 
The results using the MRST 2006 NNLO and CTEQ6.6M PDFs are quite different from 
each other at both Tevatron and LHC energies.

Ongoing work with two-loop soft anomalous dimensions in the eikonal 
approximation \cite{NKPS} and recent analytical two-loop pieces of the NNLO 
corrections \cite{NNLO2l} promise further progress in the future.

\mysection*{Acknowledgements}
 
The work of N.K. was supported by the National Science Foundation under
Grant No. PHY 0555372.  The work of R.V. was performed under the auspices 
of the U.S. Department of Energy by
Lawrence Livermore National Security, LLC, Lawrence Livermore National 
Laboratory under Contract DE-AC52-07NA27344 and also supported in 
part by the National Science Foundation Grant NSF PHY-0555660.


\begin{thebibliography}{99}

\bibitem{CDFD0}
CDF Collaboration, F. Abe {\it et al.}, Phys. Rev. Lett. {\bf 74}, 
2626 (1995) [hep-ex/9503002]; \\
D0 Collaboration, S. Abachi {\it et al.}, Phys. Rev. Lett. {\bf 74},
2632 (1995) [hep-ex/9503003]. 

\bibitem{topmass}
Tevatron Electroweak Working Group for CDF and D0, arXiv:0803.1683 [hep-ex].

\bibitem{CDFcs}
CDF Collaboration,  Phys. Rev. Lett. {\bf 96}, 202002 (2006)
[hep-ex/0603043];
Phys. Rev. D {\bf 74}, 072006 (2006) [hep-ex/0607035];
Phys. Rev. D {\bf 74}, 072005 (2006) [hep-ex/0607095];
Phys. Rev. D {\bf 76}, 072009 (2007), arXiv:0706.3790 [hep-ex].

\bibitem{D0cs}
D0 Collaboration, Phys. Rev. D {\bf 74}, 112004 (2006)
[hep-ex/0611002];
Phys. Rev. D {\bf 76}, 072007 (2007) [hep-ex/0612040];
Phys. Rev. D {\bf 76}, 092007 (2007), arXiv:0705.2788 [hep-ex];
Phys. Rev. D {\bf 76}, 052006 (2007), arXiv:0706.0458 [hep-ex];
arXiv:0803.2779 [hep-ex].

\bibitem{CDFD0st}
D0 Collaboration, V.M. Abazov {\sl et al.}, Phys. Rev. Lett. {\bf 98}, 
181802 (2007) [hep-ex/0612052];  arXiv:0803.0739 [hep-ex]; \\
CDF Collaboration, Conf. Note 8964; Conf. Note 8968.

\bibitem{topreview}
W. Wagner, Rept. Prog. Phys. {\bf 68}, 2409 (2005) [hep-ph/0507207];
A. Quadt, Eur. Phys. J. C {\bf 48}, 835 (2006);
R. Kehoe, M. Narain, and A. Kumar, Int. J. Mod. Phys. A {\bf 23}, 353 (2008), 
arXiv:0712.2733 [hep-ex];
T. Han, arXiv:0804.3178 [hep-ph];
W. Bernreuther, arXiv:0805.1333 [hep-ph].

\bibitem{NKGS}
N. Kidonakis and G. Sterman, Phys. Lett. B {\bf 387}, 867 (1996);
Nucl. Phys. {\bf B505}, 321 (1997) [hep-ph/9705234].

\bibitem{NKsingletop}
N. Kidonakis, Phys. Rev. D {\bf 74}, 114012 (2006) [hep-ph/0609287];
Phys. Rev. D {\bf 75}, 071501(R) (2007) [hep-ph/0701080]. 

\bibitem{BCMN}
R. Bonciani, S. Catani, M.L. Mangano, and P. Nason, 
Nucl. Phys. {\bf B529}, 424 (1998) [hep-ph/9801375].

\bibitem{GSWV}
G. Sterman and W. Vogelsang, JHEP {\bf 02}, 016 (2001) [hep-ph/0011289].

\bibitem{NKtop} 
N. Kidonakis, Phys. Rev. D {\bf 64}, 014009 (2001) [hep-ph/0010002].

\bibitem{KLMV}
N. Kidonakis, E. Laenen, S. Moch, and R. Vogt, 
Phys. Rev. D {\bf 64}, 114001 (2001) [hep-ph/0105041].

\bibitem{CFMNR}
M. Cacciari, S. Frixione, M.L. Mangano, P. Nason, and G. Ridolfi, 
arXiv:0804.2800 [hep-ph].

\bibitem{NKNNNLO}
N. Kidonakis, Phys. Rev. D {\bf 73}, 034001 (2006) [hep-ph/0509079].

\bibitem{NKRVtop}
N. Kidonakis and R. Vogt, Phys. Rev. D {\bf 68}, 114014 (2003) 
[hep-ph/0308222].

\bibitem{ADS}
S.M. Aybat, L.J. Dixon, and G. Sterman, Phys. Rev. Lett. {\bf 97},
072001 (2006) [hep-ph/0606254];
Phys. Rev. {\bf D74}, 074004 (2006) [hep-ph/0607309].

\bibitem{NKPS}
N. Kidonakis and P. Stephens, in {\sl DIS 2008}, arXiv:0805.1193 [hep-ph].

\bibitem{JKLT}
J. Kodaira and L. Trentadue, Phys. Lett. {\bf 112B}, 66 (1982).

\bibitem{SMPU}
S. Moch and P. Uwer, arXiv:0804.1476 [hep-ph].

\bibitem{HQFF}
W. Bernreuther, R. Bonciani, T. Gehrmann, R. Heinesch, T. Leineweber,
P. Mastrolia, and E. Remiddi, Nucl. Phys. {\bf B706}, 245 (2005)
[hep-ph/0406046].

\bibitem{NLO}
P. Nason, S. Dawson, and R.K. Ellis,
Nucl. Phys. {\bf B303}, 607 (1988);\\
W. Beenakker, H. Kuijf, W.L. van Neerven, and J. Smith,
Phys. Rev. D {\bf 40}, 54 (1989); \\
W. Beenakker, W.L. van Neerven, R. Meng, G.A. Schuler, and J. Smith,
Nucl. Phys. {\bf B351}, 507 (1991).

\bibitem{MRST2006}
A.D. Martin, W.J. Stirling, R.S. Thorne, and G. Watt, 
Phys. Lett. B {\bf 652}, 292 (2007) [arXiv:0706.0459].

\bibitem{CTEQ66}
P.M. Nadolsky, H.-L. Lai, Q.-H. Cao, J. Huston, J. Pumplin, D. Stump, W.-K. 
Tung, and C.-P. Yuan, arXiv:0802.0007 [hep-ph]. 

\bibitem{NNLO2l}
M. Czakon, A. Mitov, and S. Moch, Phys. Lett. B {\bf 651}, 147 (2007), 
arXiv:0705.1975 [hep-ph];
Nucl. Phys. {\bf B798}, 210 (2008), arXiv:0707.4139 [hep-ph];
M. Czakon,  arXiv:0803.1400 [hep-ph].

\end{thebibliography}
\end{document}